%% file: Tubiana_On_the_factorization_of_comp_knots.tex
\documentclass[notitlepage,onecolumn,amssymb]{revtex4-1} 

\usepackage{graphicx,color}
\usepackage{dcolumn}
\usepackage{bm}
\usepackage{booktabs} 
\usepackage[separate-uncertainty=true]{siunitx}  

\begin{document}

\title{
  On the factorization of composite polymer knots into separated prime
  components
}

\author{Luca Tubiana}
\affiliation{Department of Theoretical Physics, Jo\v zef Stefan Institute,
Jamova cesta 39, SI-1000 Ljubljana (Slovenia)}
\email{luca.tubiana@ijs.si}
\affiliation{
SISSA - Scuola Internazionale Superiore di Studi Avanzati Via Bonomea 265, 34136 Trieste (Italy)
}

\date{\today}

\begin{abstract}
  
  Using Monte Carlo simulations and advanced knot localization methods, we
  analyze the length and distribution of prime components in composite knots tied on
  Freely Jointed Rings (FJRs).  For increasing contour length, we observe the  progressive factorization
  of composite knots into separated prime components. However, we observe that a
  complete factorization, equivalent to the ``decorated ring'' picture, is
  not obtained even for rings of contour lengths
  $N\simeq 3N_0$, about tens of times the most probable length of the prime
  knots tied on the rings. The decorated ring hypothesis has been used in
  literature to justify the factorization of composite knots probabilities
  into the knotting probabilities of their prime components.
  Following our results, we suggest that such hypothesis may not be necessary to
  explain the factorization of the knotting probabilities, at least when
  polymers excluded volume is not relevant.  We rationalize the behavior of the system through a simple one
  dimensional model in which prime knots are replaced by sliplinks randomly
  placed on a circle, with the only constraint that the length of the loops
  has the same distribution of the length of the corresponding prime knots.
  
\end{abstract}

\maketitle

\section{Introduction}\label{sec:intro}


Knots are known to affect the physical properties of polymers~\cite{desCloizeux:1981:J-Phys-Lett,Moore:2004:PNAS, 
Stasiak:1996:Nature:8906784, Weber_et_al_2006_Biophys_J,Orlandini:2010:PRE,Saitta_et_al:1999:Nature,Arai:1999:Nature}, 
as well as the functionality of biopolymers such as DNA~\cite{Bates:DNA,meluzzi2010biophysics}. 
Knots have been found  and are actively studied in
proteins~\cite{Taylor:Nature:2000,Virnau:PLoScb:2006,King:JMB:2007,potestio:2010:PLoS,Skrbic:PLoScb:2012,Beccara:PLoScb:2013},
and in viral genomic DNA~\cite{ArsuagaPNAS2002,ArsuagaPNAS2005} where they can be used as a fingerprint to
study genome packaging in bacteriophages~\cite{MarenduzzoPNAS2009,Reith:NAR:2012}.
Furthermore, several recent studies  brought under the light the relevance of knots in nanotechnological 
applications~\cite{Arai:1999:Nature,Micheletti_PhRep2011,Micheletti:SoftMatt:2012,Rosa:PRL:2012,Coluzza:PRL:2013}.

\begin{figure}[t]
  \includegraphics[width=5.0in]{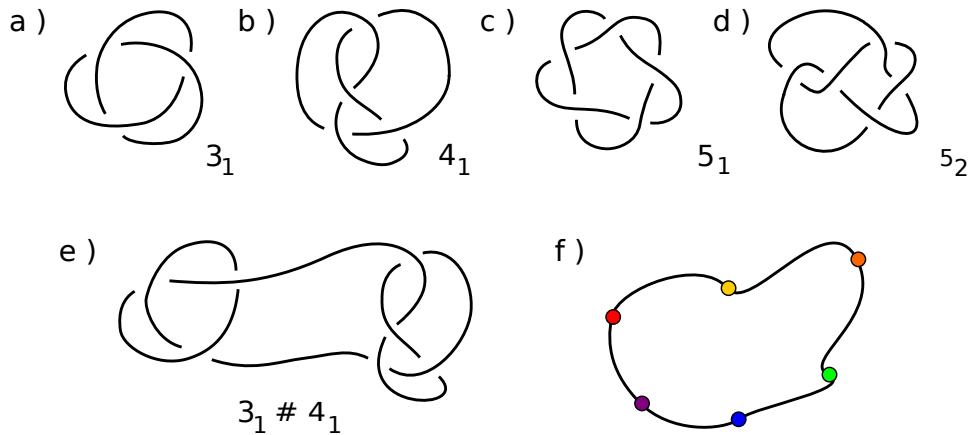}
  \caption{ The prime knot components considered in this work. a) $3_1$, 
    b) $4_1$, c) $5_1$, d) $5_2$. Panel e) shows a simple composite knot
    with two prime components, $3_14_1$ . f) In the limit of infinite length all
    component are expected to localize and behave like independent 
    point like decorations along the ring~\cite{JPA_Orlandini_1998,Baiesi:JSM:2010}.
}
\label{fig:1}
\end{figure}

Physical knots appear and diffuse spontaneously on polymer
chains~\cite{Sumners:Whittington:JPA:1988,BenNaim:prl:2001,Virnau:JACS:2005,Tubiana:Macromol:2013A}
and can be trapped by cyclization of the chains' ends. It is
possible, and frequent, for several knots to appear on the same chain, which is then said
to host a \emph{composite knot} (see Fig.~\ref{fig:1} e));  in fact,
it has been mathematically proved that such configurations are by far
the most probable for long
polymers~\cite{Sumners:Whittington:JPA:1988,Diao:JKTR:1994}.
Nonetheless, the majority of studies on the physical properties of knotted polymers still  focus on simple
(\emph{prime}) knots, while composite knots have been studied mostly in 
particular systems, like quasi-2d
rings~\cite{JPA_Guitter_1999,PhysRevLett.88.188101}, strong
polyelectrolites~\cite{Dommersnes:PRE:2002} or ring
confined into nano-channels~\cite{Micheletti:SoftMatt:2012}. In all those
systems, numerical studies shown that if several prime knots are tied on the
same chain, they do not intermingle and behave independently one from
another; composite knots are therefore ``factorized'' into
independent localized prime components and can be thought as decorated
rings, like the one depicted in Fig.~\ref{fig:1} f).  This analogy,
supported by the observed weak localization of prime
knots~\cite{PhysRevE.61.5545,Marcone:PRE:2007,Mansfield:JCP:2010}, is
invoked as well to explain the factorization of 
the knotting probability of very long three dimensional composite knotted rings into
the product of the knotting probabilities of the prime knots involved~\cite{JPSJ_Tsurusaki_1995,JPA_Orlandini_1998,Baiesi:JSM:2010}. 
 However, to the best of our knowledge, no systematic
study exists which reports on the onset of knot factorization with increasing
ring contour length; aside for the particular systems discussed above, the validity of
the assumption of complete knot factorization has not been tested.  
In this study we intend to fulfill this gap for Freely Jointed Rings,
equilateral polygon with infinitely thin edges, presenting a detailed analysis 
of the relationship between ring contour length and knot factorization. 


Using advanced knot localization algorithms~\cite{Tubiana:PTPS:2011}, we are
able to assign a knot length to composite knots, to identify all prime components 
which are separated along the ring, i.e. are not included into or
intermingled with other prime components, to measure their length,
the distribution of distances between nearest neighbors separated prime
components and the probability of having a complete factorization of composite knots
into separated prime components.

We show that even for rings whose contour length is about three times the
unknotting length, $N_0$, that is tens of times the
most probable length of the prime knots considered, the probability of still having an overlap of two or more
prime components is not negligible, going from $\simeq10\%$ for the simplest
composite knot, $3_13_1$, to $\simeq 45\%$ for the most complex composite
knot considered in this study, $3_13_13_14_1$.  Nonetheless, for contour
lengths $N\gtrsim N_0$ , the length of the composite knots as well as the
distribution of distances between nearest
neighbors separated prime components  along the ring are compatible with a random
placement of prime knots on the ring. 

Our results can be rationalized through a simple
one-dimensional model in which prime knots are substituted
by loops enforced through sliplinks. Sliplinks can be imagined as buckle
belt like rings which can move freely on the chain and enforce two
distant part of it to come near each other~\cite{Metzler:PRE:2002}.  As in
ref.~\cite{Metzler:PRE:2002} we assume that 
loops can be nested within each other, but cannot be concatenated; we further assume
that the  loops length
distributions coincide with the length distributions  of the prime knots involved in
the composite knot. We find that this simple model, with no free parameters, is able to reproduce qualitatively the
knot factorization probability curves for increasing ring contour length, while it
quantitatively reproduces the distribution of distances between separated
nearest neighbors knots and the average composite knot lengths.

The article is organized as follows. Section~\ref{sec:modmeth} reports the simulation and
analysis methods used throughout the study as well as the one-dimensional model. 
In section~\ref{sec:results} we report the results obtained from the
simulations, which are then
compared with our one-dimensional model and discussed in relation with
previous works in section~\ref{sec:disc}. 

\section{Model and methods}\label{sec:modmeth}
\subsection{The model}
\begin{figure}
  \includegraphics[width=4in]{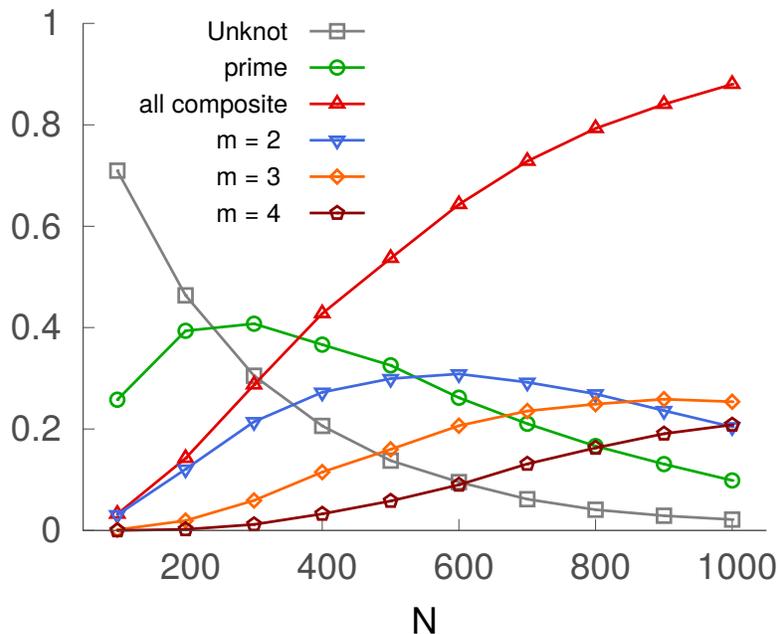}
  \caption{ Knot spectrum for Freely Jointed Rings of increasing contour length
    $N$, including unknot, prime and composite knots probabilities.
    For composite knots the plot reports the probability of a composite
    knots with an unrestricted number of components as well as those of
    composite knots with $m= 2,3$ or $4$ prime components. 
    Note that composite knots become dominant for $N \gtrsim 400$, composite
knots with $m=2$ components at $N\simeq 600$ and
composite knots with $m=3$ components at $N\simeq 900$.}
\label{fig:3}
\end{figure}

Using the crankshaft rotation algorithm~\cite{Alvarado_2011_Crankshaft},
we generated freely jointed equilateral rings of length
$N=100,200,\ldots,1000$; for comparison, the  unknotting length of FJRs is
$N_0\simeq 300$~\cite{Shimamura:JPA:2000}.  The range of lengths considered spans the crossover
from prime knots domination of the knot spectrum to composite knots
domination; composite knots start to dominate the knot spectrum for
$N\gtrsim 400$, in particular $2$-components knots start from $N\simeq 550$ and $3$-components
knots from $N\simeq 900$, as reported in Fig.~\ref{fig:3}. 

 For every length we generated $\simeq 10^7$ independent configurations with
unconstrained topology. Rings with the desired topologies were extracted from the pool of freely
jointed equilateral rings using the KNOTFIND algorithm included in the
KnotScape package~\cite{knotscape}.
We considered knots composed by $2$, $3$ and $4$ prime components, where
all but one prime components were $3_1$ knots and the remaining one
was either a $3_1$, $4_1$, $5_1$ or $5_2$ knot 
(e.g. $3_13_1$, $3_14_1$, $\ldots$, $3_13_15_1$, $3_13_15_2$,
$\ldots$).  $3_13_13_13_1$ and $3_13_13_14_1$ were the only knots with $4$
components considered in this study. For every topology and every value of $N$ we
collected several hundreds to thousands independent configurations.

\subsection{Analysis}
\begin{figure}[t]
  \includegraphics[width=4in]{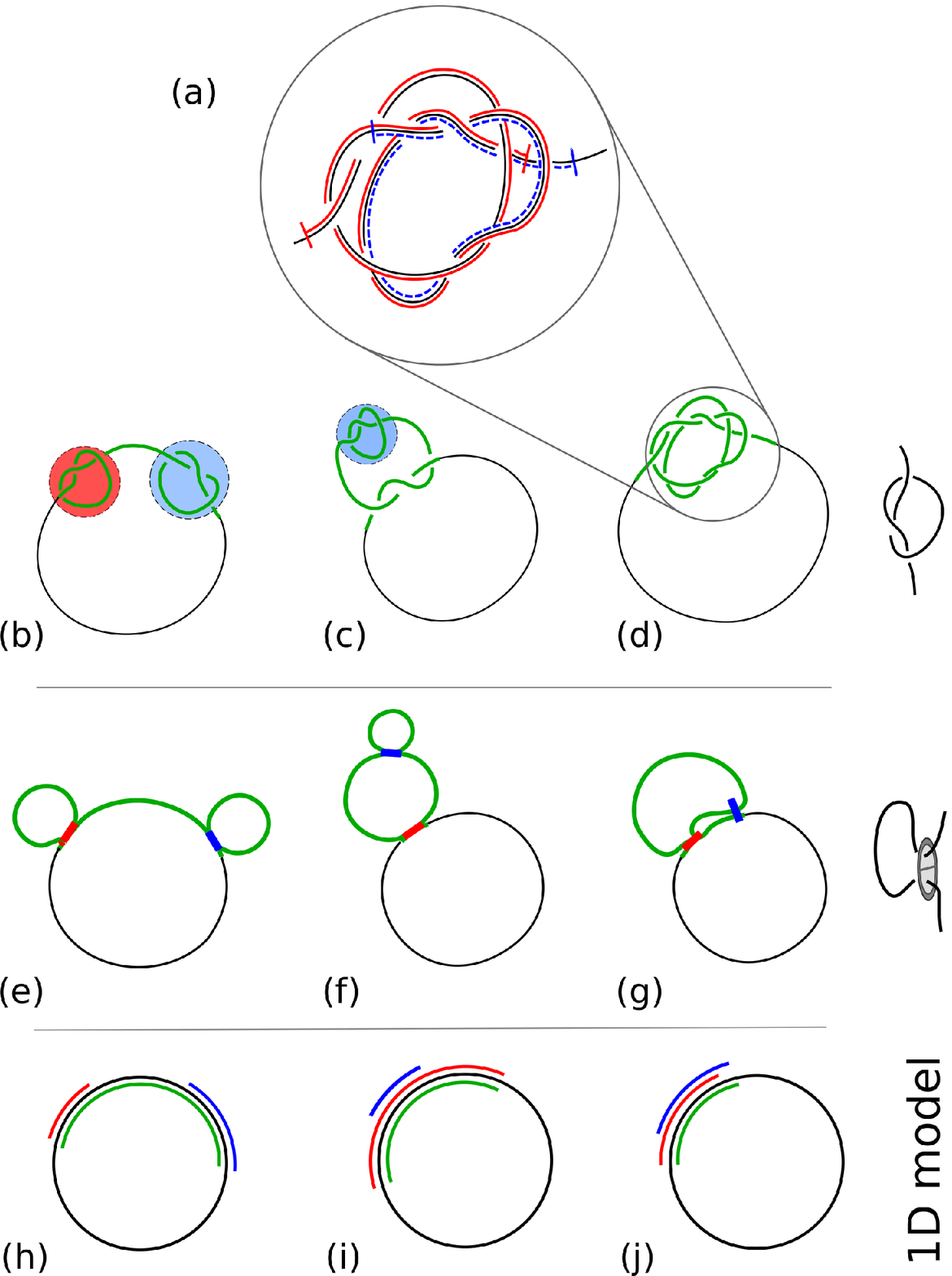}
  \caption{  a) Two prime components of a $3_13_1$ knot entangled in a way
  which does not allow the identification of a $3_1$ arc having complementary
arc of topology $3_1$. The two arcs with topology $3_1$ are coloured in
red and blue, but the condition on the complementary arc can not be
satisfied for any of them. b), c) and d) $3_13_1$ knot with $2$, $1$ and no
separated prime components respectively. 
e), f), g): sliplink representations of  knot configurations in
panels b), c), d).\\
h), i) and j): arc representations of the configurations
in b), c) and d), used to implement sliplink configurations in the
one-dimensional model. Configurations like g) and j) are
forbidden in our one-dimensional model. In  panels b) to j) the knotted portion
is highlighted in green while prime components are highlighted in red or blue.}
\label{fig:2}
\end{figure}
Consider a rope with several knots tied on it. We can intuitively say that one of
them is separated from the others if we can cut it away and rejoin the rope 
without modifying any other knot tied on it. In the pictorial examples of
Fig.~\ref{fig:2}, we can therefore find two separated components in panel b), 
one in c) and none in d).
From a mathematical point of view though, this intuitive procedure is much
subtler, as knots are defined only on closed rings; to assign a
topological state to a given ring portion, which is an open arc, one must
first close it into a ring, on which topological invariants can be
evaluated.~\cite{Millett:2005:Macromol,Virnau:JACS:2005,Orlandini:RMP:2007,Tubiana:PTPS:2011}
To close ring arcs and assign them a topological state, we adopt 
the \emph{minimally-interfering closure}, a recently introduce closure
scheme which is designed to introduce the
least possible entanglement~\cite{Tubiana:PTPS:2011}.

We consider a prime
component of topology $\tau_i$ of a composite knot of topology $\tau_{comp}
=\tau_1\ldots\tau_{i-1}\tau_i\tau_{i+1}\ldots\tau_n$ to be separated if we can 
identify, inside the composite-knotted portion of the ring,  a subarc having
topology $\tau_i$ whose complementary arc on the ring has topology
$\tau_{comp}\setminus\tau_{i}
=\tau_1\ldots\tau_{i-1}\tau_{i+1}\ldots\tau_n$. Fig \ref{fig:2} a), d) show a
pictorial example  of two prime components which are entangled to the
point that it is impossible to find any arc satisfying the above
conditions. The further condition that the prime knotted arcs must lie inside the composite-knotted
portion of the ring, i.e. inside the shortest ring portion accommodating the whole composite
knot, is set to guarantee consistency between all measured quantities. 


To identify all separate prime components, we first identified the shortest knotted portion of
the chain using a bottom-up search scheme~\cite{Tubiana:PTPS:2011}, which
identifies the shortest arc of the chain which has the desired topology
along with an unknotted complementary arc. Then 
we ordered all the subarcs of the shortest knotted portion for increasing
length. Starting from the shortest arc we checked if it was a factor prime knot 
of the composite knot. If this was the case, all the arcs overlapping it were 
removed from the list of arcs to be analyzed. This procedure guarantees that in a situation like that depicted in
Fig.~\ref{fig:2} c)  only one component is seen as separated, while
the larger knotted arc is discarded as it includes the smaller one and therefore
can not be removed from the ring without removing the other knot along with
it. 

With the above procedure we computed the number of separated prime components
in each configuration; the probability of having all prime components
separated on a ring of length $N$,
$P_{sep}(\tau_{comp},N)$, estimated as the ratio between the number of configurations of 
topology $\tau_{comp}$ in which all prime components have been identified and the
total number of configurations of that topology; the average length of
separated prime components knots, $\langle l_k \rangle$, and the length of the composite knot, $\langle
l_k^{comp} \rangle$.



\subsubsection{Ring simplification}
In order to reduce both the computational cost of localizing
the knots and the incidence of slipknots~\cite{Millett:JKTRL:2010,Tubiana:PTPS:2011} 
we applied a topology-preserving simplification procedure. 

The adopted simplification procedure is an improved version of the
rectification scheme adopted in~\cite{Koniaris:prl:1991,JPSJ_Tsurusaki_1995,Taylor:Nature:2000}.
The standard rectification scheme consists in picking a random vertex $i$ of
the ring and verify if it can be made collinear with its neighbor
vertices $i-1$ and $i+1$ without crossing any edges of the ring. If it is
possible, the vertex is eliminated. The procedure is then carried on until
no more vertices can be eliminated. This simplification scheme alters the
geometric properties of the ring, and can therefore modify some geometrical
properties of the knots. In order to reduce these effects, we
imposed a \emph{simplification stride}, $s$, that limited the number of subsequent
vertices that could be eliminated. We started with $s=2$ and applied the
rectification scheme $N$ times, rejecting those vertex removals which would introduce a 
gap larger than $s$ between the remaining beads. We then repeated this procedure
with a new value of $s$, $s' = 2*s$ until we reached a maximum value of stride
$s_{max} = 15$ which was chosen for it being close to the
characteristic length of trefoil knots on FJRs (see supplementary
information).

To obtain an estimation of the error introduced by the simplification
procedure, we repeated the analysis on small sets of hundreds of
configurations for several topologies for two other values of $s_{max}$: $s_{max} = 4$
and $s_{max}=\infty$. We observed that varying $s_{max}$ the length of the
composite knots remained compatible within the errors, while $P_{sep}$
diminished systematically for increasing $s_{max}$, with a maximum
difference in the estimate of $P_{sep}(3_13_13_13_1)$ of $\simeq 10\%$
measured between $s_{max}=4$ and $s_{max}=\infty$. See supplementary
information for further details.

\subsection{One-dimensional model}\label{sec:method1D}

To gain further insight into the behavior of composite knots, we designed a
simple one-dimensional model in which prime factor knots are substituted with
loops enforced by sliplinks~\cite{Metzler:PRE:2002}
placed randomly on a plane, equilateral, polygon with $N$ edges, as
illustrated in Fig.~\ref{fig:2} e)-j).
Any two loops on the polygon can only be either separated or nested
within each other, so configurations like that of Fig.~\ref{fig:2} g) and j) are
prohibited. 
The length distribution of the loops enforced by the sliplinks is assumed to be that of prime
knots of the same topology. Following ref.~\cite{Metzler:PRE:2002} we call these structures 
formed by sliplinks on a ring ``paraknots''.

As $3_1$ and $4_1$ prime knots are those for which we have more statistics, 
we considered the following topologies: $3_13_1$, $3_14_1$, $3_13_13_1$, $3_13_14_1$,
$3_13_13_13_1$ and $3_13_13_14_1$. The length distribution of the loops is
given by the empirical knot length distributions, $P(l_k,N)$,
of prime knots $3_1$ and $4_1$ extracted from the pool of FJRs, evaluated
with a bottom-up scheme (See Supplementary material). We stress that these are  knot length
distributions of prime knots, and not of prime components in a composite knot.

To generate paraknots with $m$ loops on polygons with $N$ edges, we 
first picked randomly $m$ edges of the polygon, 
corresponding to the starting points of the arcs representing the loops (in a clockwise
direction). To every starting point we then associated a knot length picked randomly
from the length distribution $P(l_k,N)$ of the corresponding prime knot
($3_1$ or $4_1$) and computed the ending points of the arcs, taking into
account the periodicity of the polygon. All configurations in which
two or more arcs overlapped without one being strictly included into the other
were rejected  (e.g. Fig.~\ref{fig:2} j)).

For each topology and ring length we generated several thousands paraknots configurations. On these configurations we
performed the same analysis as on the configurations from the simulation.
In this model, we defined the knotted portion of the ``composite knot'' 
as the shortest arc of the polygon which includes all the sliplinks, i.e. all
the starting and ending points. Since a polygon with $m$
sliplinks is divided in $2m$ ``unknotted'' arcs, the ``knotted portion''
is the complementary of the longest unknotted arc.

\section{Results}\label{sec:results}

\subsection{Knot factorization probability}

\begin{figure}[t!]
\includegraphics[width=4in]{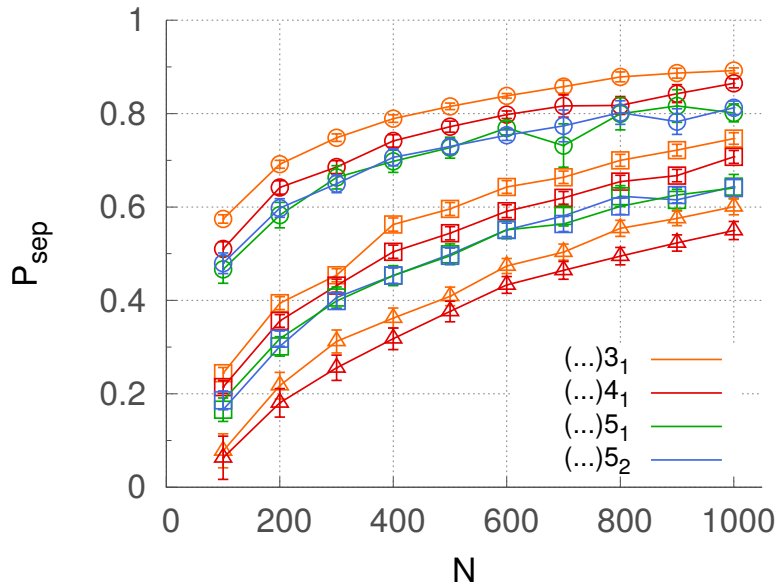}
\caption{
  Probability, $P_{seq}(\tau,N)$ of having all prime components separated,
  plotted as a function of ring length $N$, for different composite knots.
  From top to bottom, the lines report the probabilities for knots
with $2$, $3$ and $4$ components with topologies  ($\ldots)3_1$, ($\ldots)4_1$,
($\ldots)5_1$, ($\ldots)5_2$ respectively.}
\label{fig:4}
\end{figure}

To investigate whether and to what extent composite knots factorize into
separated prime components with increasing ring contour length $N$, we study the probability,
$P_{sep}( \tau_{comp}, N)$, of having all prime components
of a composite knot of topology $\tau_{comp}$ separated along the chain. 
The results are plot in Fig.~\ref{fig:4}, and show that while the probability of
having all factor knots separated increases with $N$, a complete factorization
is not reached in the range of lengths investigated. Indeed, even for the
simplest topology, $3_13_1$, and the longest length, $N=1000$, which is about
$100$ times the typical length of a trefoil knot (see Supplementary
material), the probability of having at least one non-separated component,
$1-P_{sep}(\tau_{comp},N)$, is still $~\simeq 10\%$.  This is arguably due to the fact that
knot length distributions of prime knots, although being peaked at low
values of $l_k$ independently of $N$~\cite{Mansfield:JCP:2010,Tubiana:Macromol:2013A}
( $l_k\simeq 10$ for $3_1$ knots, see supplementary material), have very long tails which give origin to the sublinear
growth of the average knot length with $N$ (weak knot localization)~\cite{Marcone:PRE:2007}. Furthermore, observing
the slopes of the $P_{sep}(\tau_{comp},N)$ curves, it is easily seen that not only a
complete factorization is not reached in the analyzed length range, which
reaches  $N \gtrsim 3 N_0$, but it should not be expected even for far
larger rings.

Comparing the results for knots with $2,3$ or $4$ prime components, we
observe that $P_{sep}(\tau_{comp},N)$ depends on the number of knots present and their
minimum crossing number. 
The behavior of knots with either a $5_1$ or
a $5_2$ prime component is identical, suggesting that, at least for simple
prime factor knots, the only discriminant is the minimum crossing number.

\subsection{Knot lengths}

To better characterize the behavior of composite knots, we turn to study the average
lengths of whole composite knots, $\langle l_{k}^{comp} \rangle$, and of
their separated
prime components, $\langle l_k \rangle$, as a function of both the contour length of the ring and 
the number of prime components of the knot.
\begin{figure}
\includegraphics[width=4in]{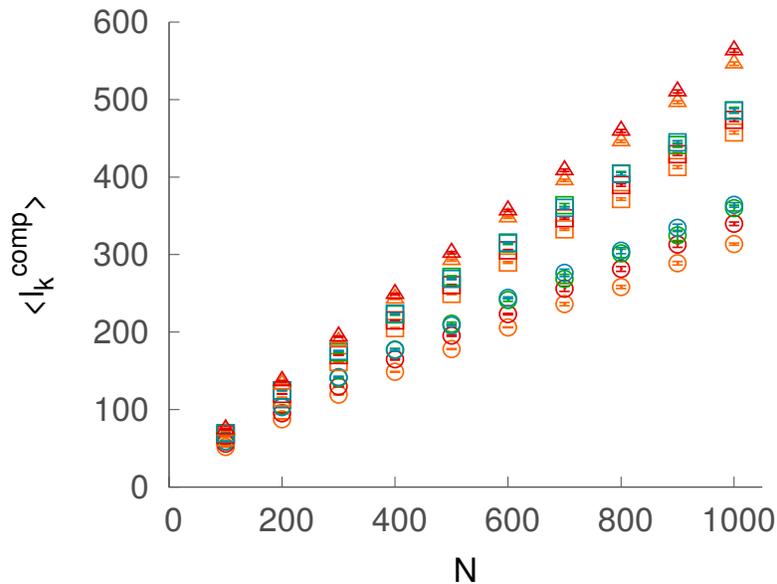}
\caption{ From top to bottom, average lengths, $\langle
    l_k^{comp}\rangle$, of composite knots with
    $4$, $3$ and $2$ components. Colors and symbols follow the scheme
    introduced in Fig.~\ref{fig:4}.}
\label{fig:5}
\end{figure}
As reported in Fig.~\ref{fig:5}, we observe that
$\langle l_k^{comp}\rangle$ grows linearly in $N$, compatibly with the
factor knots being randomly distributed on the ring. Fitting the lines for
$N\geq 500$, we obtain the slopes reported in table~\ref{tab:knotlengths}.
It is clear from the reported values that the slopes depend mainly on
the number of prime components forming the knot and only
slightly on the exact topology of the factor knots. 
As a comparison, we consider the value of the knot length which would be
obtained if the knots were equivalent to point-like decorations along the
rings. In this limit case, $l_k^{comp}$ is given by the length of the
shortest arc hosting all the decorations and  depends only on
the number of point-like decorations present on the ring.  The expectation value of the knot
length in this limit case can be computed analytically and is $\bar{l}_{2 points}= 1/4 L$, $\bar{l}_{3 points} = 7/18
L$ and $\bar{l}_{4 points} = 23/48 L$ for two, three and four points placed randomly on a circle
of contour length $L$. The slopes obtained from our simulations are
larger than those factors, showing that finite size effects are still
important in the range of ring lengths considered.
\begin{figure}
\includegraphics[width=4in]{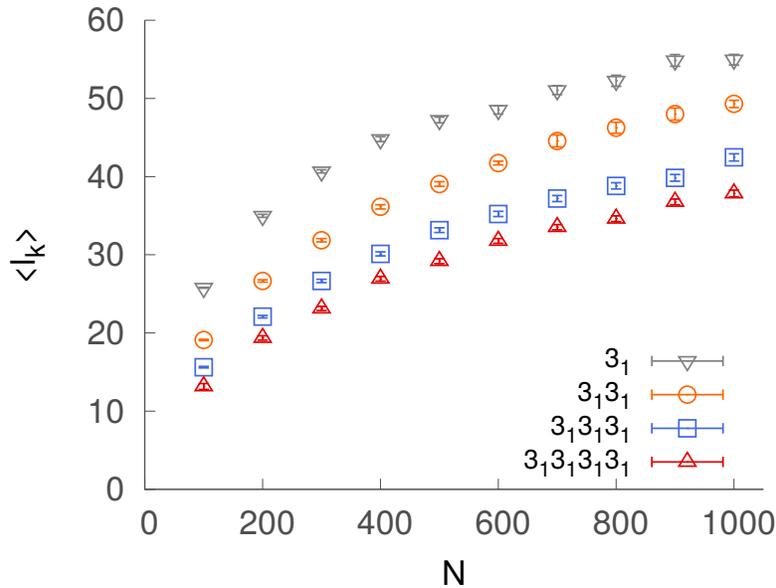}
\caption{ From top to bottom, average knot lengths, $\langle l_k\rangle$, of prime
    trefoil knots and of trefoil factor knots in composite knots with $2$, $3$ and $4$
    components. The average is computed only on separated components.
    }
\label{fig:6}
\end{figure}

Finally, looking at $\langle l_k\rangle$, reported in
Fig.~\ref{fig:6}, we observe that the size of separated 
prime components is influenced by the number of prime knots tied on the ring,
another consequence of the fact that knot localization is only weak.
Even the average lengths of separated prime components in a
$3_13_1$ knot is still lower than that of a prime $3_1$ knot, while in the
asymptotic limit one would expect the lengths of prime factor knots to be
independent of their number.

\subsection{Factor knots placement on the ring}

The linear growth of  $\langle l_k^{comp}\rangle$ with $N$ is compatible with factor knots being
randomly distributed along the ring. Nonetheless, there is still the possibility that 
two factor knots interact in some way when they are close to each other,
for example by exchanging knot length or by slightly repelling each other
because of an entropic interactions of their loops.  
If this is the case, we expect that some effects will be visible in the
distribution of factor knot lengths, $l_k$,  at different separations of two factor
knots along the chain.  

To investigate whether factor knots interact, we consider
only those configurations in which all factor knots are separate. 
Going clockwise along the ring, for every couple of adjacent 
factor knots we compute their separation along the ring, $s^{AB}$, from the
ending point of one factor knot 
to the starting point of the next one, and the sum of their knot lengths
$l_k^{AB} = l_k^A+l_k^B$. We then average all lengths $l_k^{AB}$ over bins of length $10$ 
in $s^{AB}$ and over the whole set of configurations, 
to observe if the average knot length of two adjacent components
depends on their separation. 

\begin{figure}
\includegraphics[width=4in]{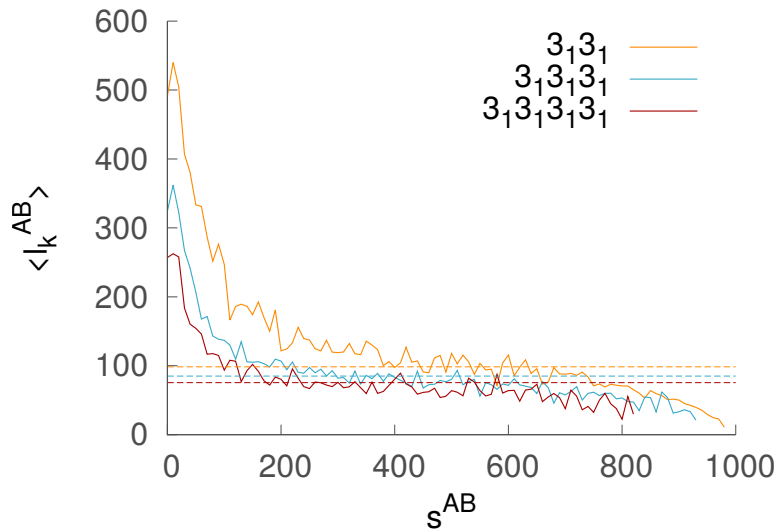}
\caption{Average total length of two factor knots, $<l_k^{AB}>$, as a function of their
separation, $s^{AB}$, along the ring, for three different composite knots on
rings with $N=1000$ edges. Horizontal dashed lines correspond to the values
$2*\langle l_k\rangle$ reported in Fig.~\ref{fig:6}}
\label{fig:7}
\end{figure}

The results are shown in Fig.~\ref{fig:7}  for rings of length 
$N=1000$. One can see that going from small to 
large separations, the length $<l_k^{AB}>$ has a pronounced peak at
first, followed by a mild decrease which becomes more pronounced when
$s^{AB}\to N$. 

The behavior of $\langle l_k^{AB}\rangle$ may be explained without
accounting for an interaction between factor knots, by considering that, since the
contour length $N$ is fixed, large factor knots are forced to stay closer to
each other than small ones.
\begin{table}
  \begin{tabular}{ccS[table-format=1.3(3)]ccS[table-format=1.3(3)]ccS[table-format=1.3(3)]}
  \toprule
  Topology&{ }&{ }&{ }&{ }&{ Number of Components }&{ }&{ }&{ }\\
  { }&{ }&{2 }&{ }&{ }&{3 }&{ }&{ }&{4 }\\
  \midrule
  $(\ldots)3_1$ &{ }& 0.271\pm0.005&{ }&{ }& 0.414\pm0.003&{ }&{ }& 0.504\pm0.004\\
  $(\ldots)4_1$ &{ }& 0.290\pm0.004&{ }&{ }& 0.423\pm0.003&{ }&{ }& 0.519\pm0.003\\
  $(\ldots)5_1$ &{ }& 0.294\pm0.006&{ }&{ }& 0.427\pm0.008&{ }&{ }& { -- }\\
  $(\ldots)5_2$ &{ }& 0.308\pm0.006&{ }&{ }& 0.433\pm0.006&{ }&{ }& { -- }\\
\bottomrule\\
\end{tabular}
\begin{flushleft}
  \caption{ Fits obtained by linear regression of the data reported in
    Fig.~\ref{fig:5}, for $N\geq500$. 
    The correlation coefficient is $\simeq 0.999$ for all fitting
lines.}
\label{tab:knotlengths}
\end{flushleft}
\end{table}

\section{Discussion}\label{sec:disc}

The results of Sec.~\ref{sec:results} show that the paradigm of prime
components behaving as decorations on the ring does not hold for FJRs in the range of length considered; nonetheless 
prime knots appear to be distributed randomly on the ring. Those results can
be interpreted in the light of a transparent one-dimensional model, based on
the following assumptions.
\begin{figure}
\includegraphics[width=7in]{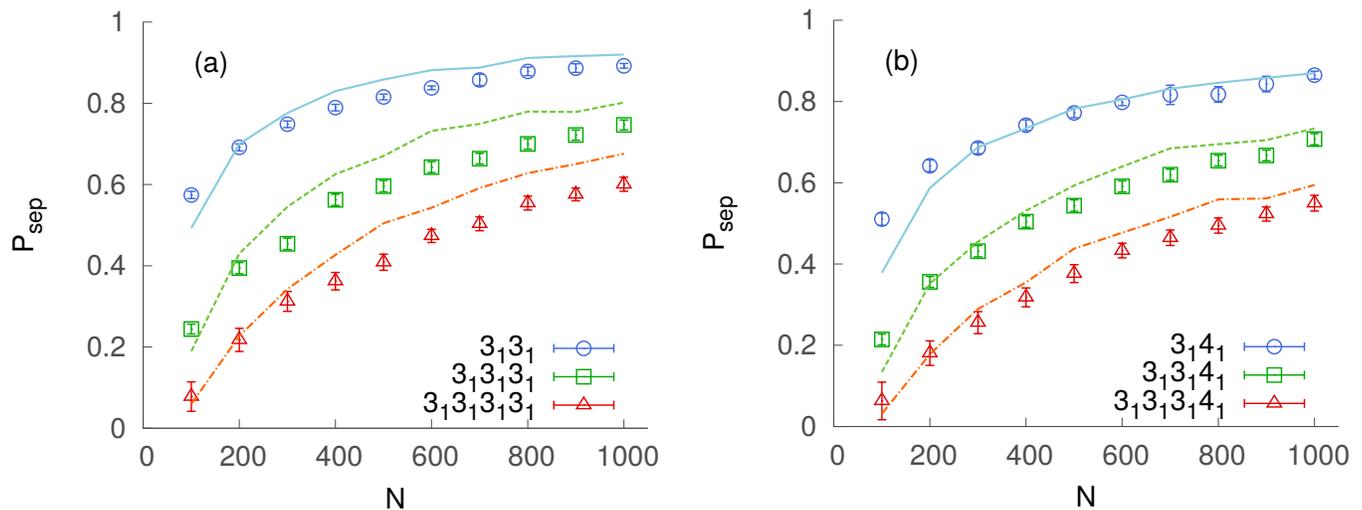}
\caption{Comparison between the probabilities $P_{sep}(\tau_{comp},N)$ obtained from the
  simulations (points) and from the 1D model (lines) for composite knots
made only of $3_1$ knots (a) and composite knots including a $4_1$
factor (b). }
\label{fig:8}
\end{figure}
\begin{figure*}[t]
\includegraphics[width=7in]{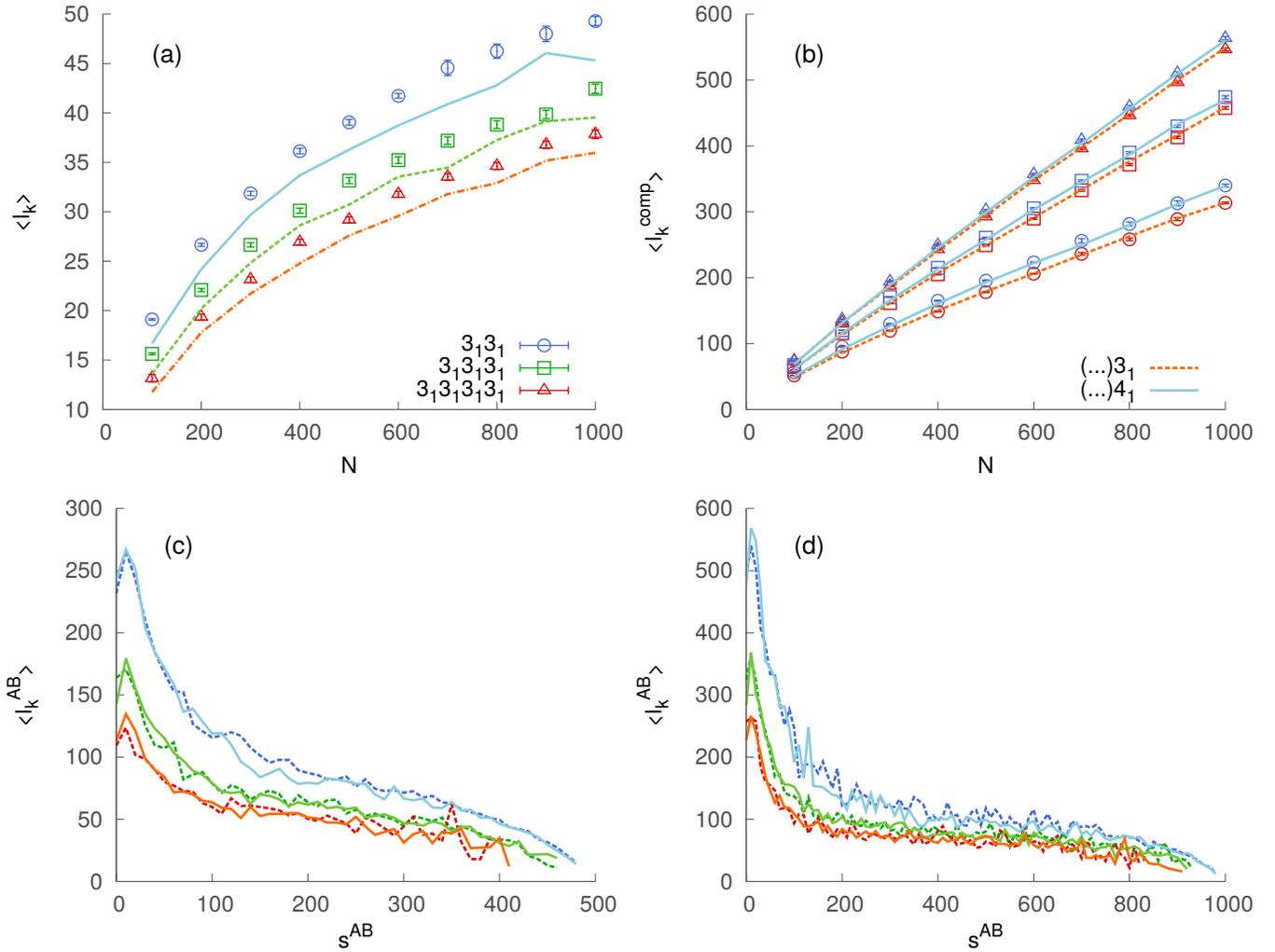}
\caption{ (a) Length of factor knots $l_k$. (b) Total length of composite
knots, $\langle l_k^{comp}\rangle $. Comparison between simulation and 1D model for the distribution of
total knot length $\langle l_k^{AB}\rangle $ of couples of adjacent factor knots separate
by a distance $s^{AB}$ along the ring, for rings of length $N=500$ (c) and
$N=1000$ (d). Green, blue and red dashed lines report simulation data for  
$3_13_1$,  $3_13_13_1$, $3_13_13_13_1$ topologies. Light green, cyan and orange
lines report data for the same topologies obtained from the model.}
\label{fig:9}
\end{figure*}

\begin{enumerate}
  \item All factor knots are placed randomly on the ring.
  \item Two factor knots can only be either separated along the chain or
    nested, see Fig.~\ref{fig:2} b), c). This assumption lies on the observation that  
    entangled configurations like the one depicted in Fig.~\ref{fig:2} a),d) are very rare 
    (less than $1\%$ for two components knots and $N\geq400$) and can, to a first approximation, be neglected. 
  \item The probability that a factor knot  $\tau_i$ has length $l_k$ 
    depends on the presence of other knots only through the imposition of
    the previous condition. In sufficiently long rings, the length distribution of factor knots 
    can be considered to be practically identical to that of their corresponding prime knots.
\end{enumerate}
According to the first and second assumptions, we describe factor knots as
paraknots placed randomly on a ring.
 Each paraknot is identified by a sliplink placed on the ring, joining
two distant vertices of it. Every sliplink identifies a loop.  No sliplink
can be placed so to join a point inside a paraknot with one outside it: configurations like that
depicted in Fig.~\ref{fig:2} g), j) are forbidden. According to our third hypothesis, we impose the
length distribution of the paraknots to be the same as that of the prime
knots they model.  

Following the procedure reported in \S\ref{sec:method1D}, we repeated the
analyses of Sec. \ref{sec:results} for the one-dimensional model.
A direct comparison of the various quantities obtained
from the one-dimensional model with those obtained on FJRs is reported in
Fig.~\ref{fig:8} and Fig.~\ref{fig:9}. Results from the one-dimensional model are in
qualitative agreement with those obtained from the simulations for the
probability of having all prime components separated, $P_{sep}(\tau_{comp},N)$, reported in
Fig.~\ref{fig:8} a), b), and for the behavior of $\langle l_k\rangle$
with $N$ and $m$, reported in Fig.~\ref{fig:9} a).

The agreement between the one-dimensional model and the data from
simulations becomes quantitative when we take into account more global
properties, like the length of the composite knot as a whole, $\langle
l_k^{comp}\rangle$, and the distribution of  $\langle l_k^{AB}\rangle $ as a function of the linear separation between
adjacent couples of factor knots, $s_{AB}$, reported in Fig.~\ref{fig:9}
b), c) and d). 

The discrepancies observed between the one-dimensional model and the
simulations data can be ascribed to the simplification procedure adopted to
measure knot lengths in the simulations. The simplification procedure results 
usually in a slight overestimation of the lengths of knots. Consequently,
two factor knots which are  very close to each other along the
chain may be seen as not being separate, lowering $P_{sep}$.
Indeed, while in the case of $P_{sep}$ the
quantities obtained from the model are systematically higher than those
obtained from simulations, in the case of the length of factor knots they are
systematically lower. The underestimation is expected to increase for increasing number of components, as
observed in Fig.~\ref{fig:8}. 

It is interesting to compare our results against previous results on the factorization
of knotting probability obtained in ref.~\cite{JPSJ_Tsurusaki_1995} for Gaussian
random polygons and in ref.~\cite{Baiesi:JSM:2010}
for self-avoiding rings on the cubic lattice. In both cases the 
factorization was shown to hold for lengths in the order of $\simeq N_0$. 
The results reported in Sec.~\ref{sec:results}, in which no complete
factorization into separated prime components was observed, and their agreement with
the model presented in this section, raise the question if the ``decorated
ring'' assumption~\cite{JPSJ_Tsurusaki_1995,JPA_Orlandini_1998,Baiesi:JSM:2010}
is really necessary to guarantee  the factorization of the knotting
probability of composite knots into the product of those of their prime components, or if a
mechanism like the one introduced by our one-dimensional model will be
sufficient.

It is also interesting to consider what happens when one
studies an equilibrium population of composite knots obtained from ring polymers circularized at $\theta$-point. 
Our results show that the probability $P_{sep}(\tau_{comp},N)$ decreases with the number of prime 
components tied on a ring of length $N$. Since for increasing ring length the
average number of prime components involved in a composite knot increases
(see Fig.~\ref{fig:3}), it is possible that in such sample no complete
factorization into separated prime components will be observed, on average, at any ring contour length.

\section{Summary and Conclusions}

We have presented a numerical study of the properties of composite knots
made of up to $4$ components. Using advanced knot localization methods, we
have been able to identify separated prime components, measure their length and
separation, as well as the probability of having a full factorization of
composite knots into separated prime components.

We have shown that a complete factorization of a composite knot
into separate prime components is not to be expected in the range of
lengths considered; nonetheless the characteristics of composite knots are 
compatible with a random placement of their prime components along the
ring. We rationalized the simulation results using 
a simple one-dimensional model, in which factor knots are substituted with
randomly placed paraknots whose length distribution coincide with that of
the factor knots.

Interestingly, the range of ring contour lengths
considered in this study is comparable, in relation with the unknotting
length, to those considered in previous studies~\cite{JPSJ_Tsurusaki_1995,Baiesi:JSM:2010},
in which the factorization of the knotting probability of composite knots
was observed in two different polymer systems: Gaussian
polygons~\cite{JPSJ_Tsurusaki_1995} ( no excluded volume) and self-avoiding rings on the square
lattice~\cite{Baiesi:JSM:2010}.  Following this observation we suggest that,
at least for polymer models without excluded volume, the ``decorated ring''
paradigm  is not necessary to explain the factorization of knot
probabilities, and may be substituted by the ``paraknot'' paradigm used to
explain our data.  Since our study was performed on infinitely
thin polymer rings, it will be interesting to investigate whether the introduction
of excluded volume leads to a faster knot factorization, and thus to the
``decorated ring'' paradigm, or if the behavior in the presence of
excluded volume is still compatible with the one reported here.  

Finally, it is known that geometrical confinement affects knot properties, including knot
localization~\cite{Tubiana:PRL:2011,Micheletti:Macromol:2012,Micheletti:SoftMatt:2012}. In
particular, prime knots confined into  nano-channels~\cite{Micheletti:SoftMatt:2012} and quasi 
two dimensional knots~\cite{JPA_Guitter_1999,PhysRevLett.88.188101} are much
more localized that their unconstrained counterparts studied here. An higher
degree of localization intuitively leads to a more effective factorization of composite knots into
separated prime components and, in fact, in such cases, composite knots have been shown to be
completely factorized, and therefore  equivalent to decorated rings~\cite{PhysRevLett.88.188101,Micheletti:SoftMatt:2012}. 
We think it will be interesting to extend the present work to study the
progressive onset of knot factorization as a function of increasing geometrical
confinement.

\section{Acknowledgements}
I am deeply indebted to C. Micheletti for his support in this project. I am
also very grateful to E. Orlandini and E. Paoli for several useful discussions.
I acknowledge financial support from the Slovene Agency for Research and
Development ( Grant No. J1-4134).

%

\newpage
\section*{Supplementary Information}
\input{Tubiana_On_the_factorization_of_comp_knots_SUPP_MAT.tex}
\end{document}

%% file: Tubiana_On_the_factorization_of_comp_knots_SUPP_MAT.tex
\setcounter{figure}{0}

\renewcommand{\figurename}{SUPPL. FIG.}
\begin{figure*}[h]
\includegraphics{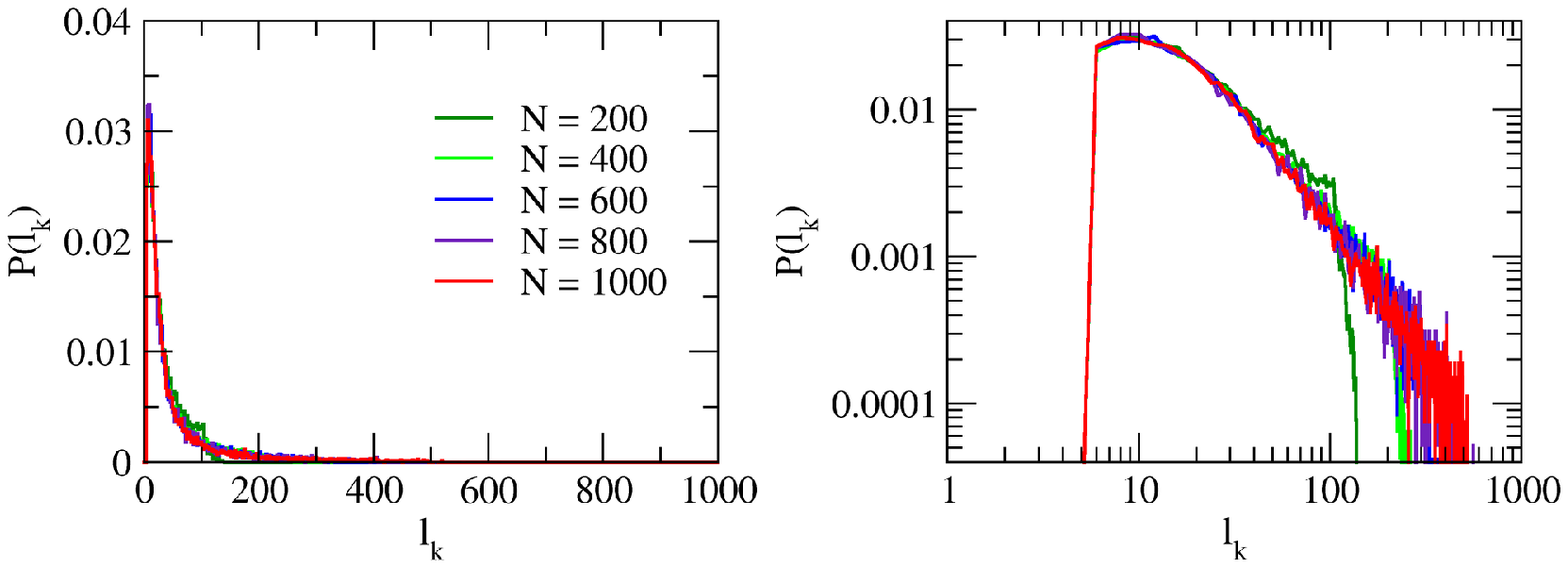}
\caption{Knot length distribution for trefoil knots tied 
  on rings of increasing length $N$. Note that all distributions are peaked on the same value
$l_k\simeq 10$ while they differ because of their tails.}
\label{fig:SI1}
\end{figure*}

\begin{figure*}
\includegraphics[width=5.0in]{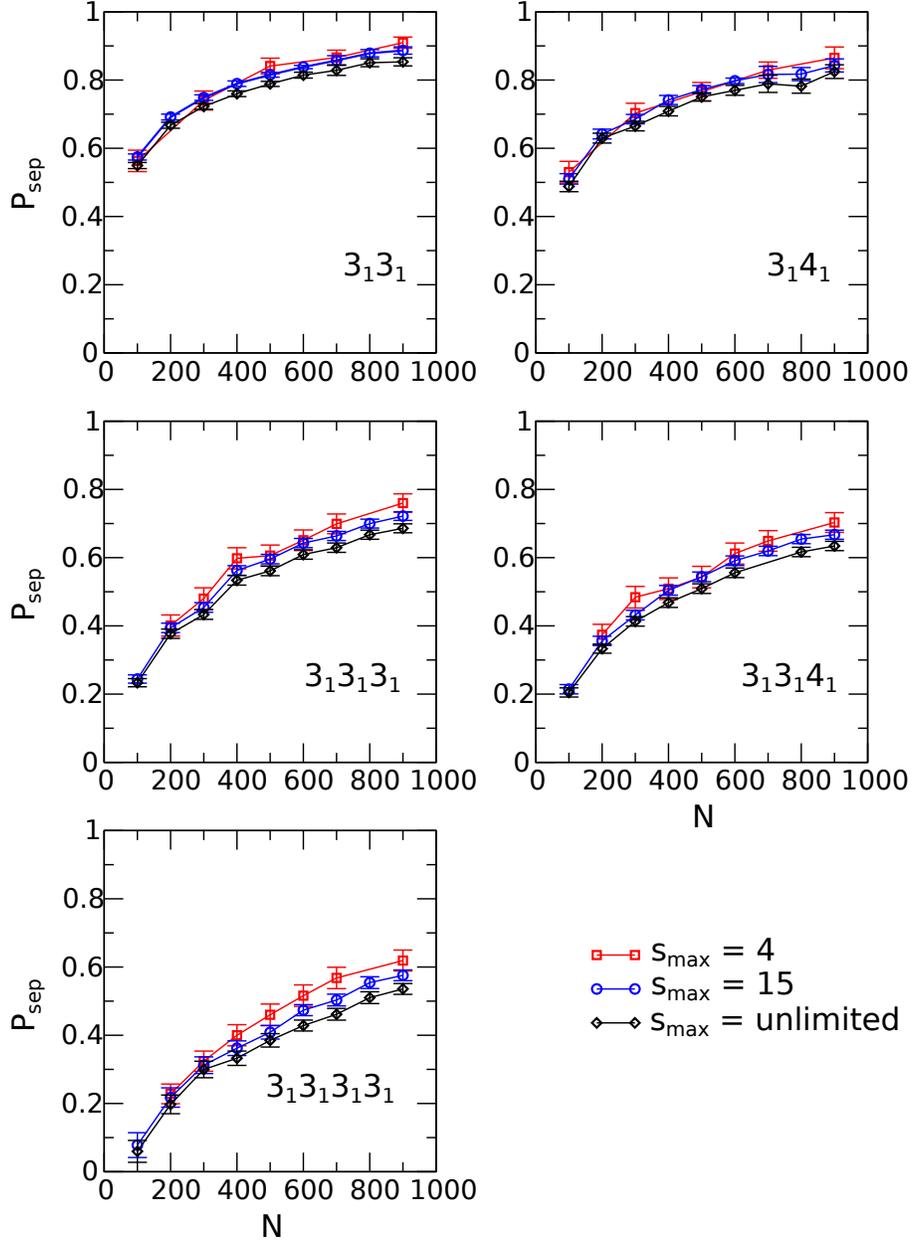}
  \caption{Different measures of $P_{sep}$ done on smaller sets of
  configurations with different simplification strides. The difference
between the probabilities measured with different strides increases with the
number of factor knots, but remains between  $\simeq10\%$.}
\label{fig:SI2}
\end{figure*}